\documentclass[prr,superscriptaddress,longbibliography,twocolumn]{revtex4-1}
\usepackage[utf8]{inputenc}
\usepackage[english]{babel}
\usepackage{amsmath}
\usepackage{amsfonts}
\usepackage{amssymb}
\usepackage{graphicx}
\usepackage{amsthm}
\usepackage{color}
\usepackage{hyperref}
\usepackage[caption=false]{subfig}
\usepackage{graphicx,float}
\usepackage{dcolumn}
\usepackage{bm}
\usepackage{mathrsfs}
\usepackage{txfonts}
\usepackage{CJK}
\usepackage[amssymb]{SIunits}
\usepackage{epsfig}
\usepackage{epstopdf}
\usepackage{lipsum}
\usepackage{placeins}
\usepackage{blkarray}
\newcommand{\ket}[1]{\ensuremath{|#1\rangle}}
\newcommand{\bra}[1]{\ensuremath{\langle #1|}}

\newenvironment{SChinese}{%
	\CJKfamily{gbsn}%
	\CJKtilde
	\CJKnospace}{}
\allowdisplaybreaks[2]

\begin{document}

\begin{CJK}{UTF8}{}
\begin{SChinese}

\title{High-Dimensional Two-Photon Quantum Controlled Phase-Flip Gate}

 \author{Mingyuan Chen}  %
 \affiliation{College of Engineering and Applied Sciences, National Laboratory of Solid State Microstructures, Nanjing University, Nanjing 210023, China}

\author{Jiang-Shan Tang}  %
\affiliation{College of Engineering and Applied Sciences, National Laboratory of Solid State Microstructures, Nanjing University, Nanjing 210023, China}

\author{Miao Cai}  %
\affiliation{College of Engineering and Applied Sciences, National Laboratory of Solid State Microstructures, Nanjing University, Nanjing 210023, China}

\author{Yanqing Lu}  %
\email{yqlu@nju.edu.cn}
    \affiliation{College of Engineering and Applied Sciences, National Laboratory of Solid State Microstructures, Nanjing University, Nanjing 210023, China}
 
\author{Franco Nori (野理)}
 \email{fnori@riken.jp}
\affiliation{Quantum Computing Center,  Cluster for Pioneering Research, RIKEN, Wako-shi, Saitama 351-0198, Japan}
\affiliation{Physics Department, The University of Michigan, Ann Arbor, Michigan 48109-1040, USA}

\author{Keyu Xia (夏可宇)}  %
 \email{keyu.xia@nju.edu.cn}
    \affiliation{College of Engineering and Applied Sciences, National Laboratory of Solid State Microstructures, Nanjing University, Nanjing 210023, China}
  \affiliation{Hefei National Laboratory, Hefei 230088, China}

\date{\today}

\begin{abstract}
High-dimensional quantum systems have been used to reveal interesting fundamental physics and to improve information capacity and noise resilience in quantum information processing. However, it remains a significant challenge to realize universal two-photon quantum gates in high dimensions with high success probability. Here, by considering an ion-cavity QED system, we theoretically propose, to the best of our knowledge, the first high-dimensional, deterministic and universal two-photon quantum gate. By using an optical cavity embedded with a single trapped $^{40}\textrm{Ca}^{+}$ ion, we achieve a high average fidelity larger than 98\% for a quantum controlled phase-flip gate in four-dimensional space, spanned by photonic spin angular momenta and orbital angular momenta. Our proposed system can be an essential building block for high-dimensional quantum information processing, and also provides a platform for studying high-dimensional cavity QED.
\end{abstract}

\maketitle

\end{SChinese}
\end{CJK}


\begin{figure}
  \centering
  \includegraphics[width=1.0\linewidth]{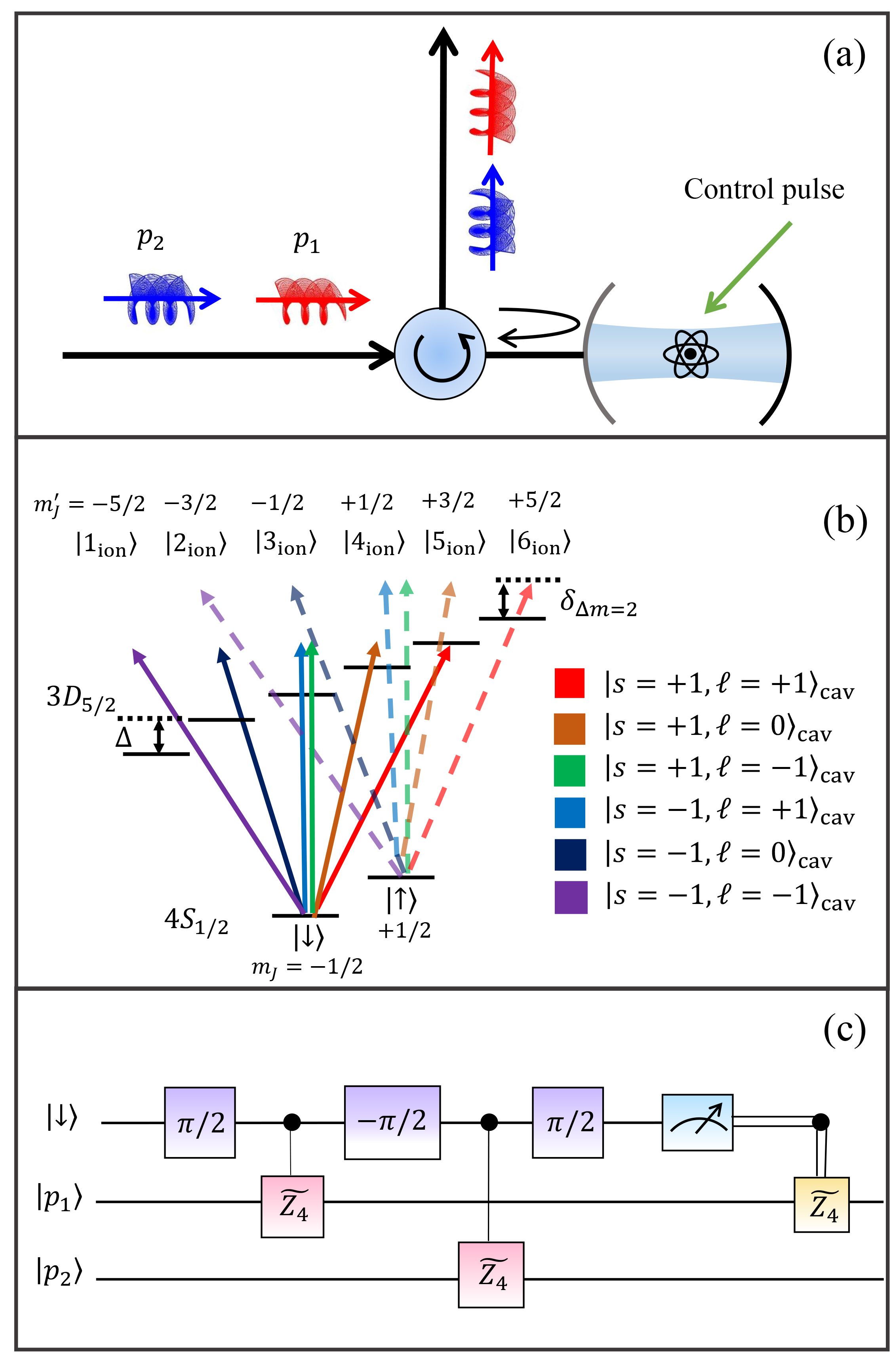} \\
\caption{Schematics of the high-dimensional two-photon quantum controlled phase-flip gate. (a) Two single-photon pulses $p_1$ and $p_2$ carrying SAM and OAM act as 4D qudits. They enter the Fabry-P\'{e}rot cavity successively as two separate spatio-temporal modes via an optical circulator. The photons are subsequently reflected from the cavity containing a single $^{40}\text{Ca}^{+}$ ion and acquire a correlated $\pi$ phase shift. (b) Involved energy levels of the $^{40}\text{Ca}^{+}$ ion. Transitions are driven by photons in different combinations of SAM and OAM. Transitions between the state $\mathinner{|\!\uparrow\rangle}$ and the excited magnetic sublevels, $\{|m^\prime_J = -3/2\rangle, |m^\prime_J = -1/2\rangle, |m^\prime_J = 1/2\rangle, |m^\prime_J = 3/2\rangle, |m^\prime_J = 5/2\rangle\}$, are far off resonance and negligible. (c) Quantum circuit showing steps performing the proposed quantum gate.}
\label{fig:FIG1}
\end{figure}

\section{Introduction}
Cavity and waveguide quantum electrodynamics (QED) systems have demonstrated the powerful capability of controlling transport of photons by exploiting the strong interaction between atoms and photons in an optical cavity or a waveguide~\cite{ RevModPhys.94.041003}, both theoretically~\cite{ PhysRevLett.92.127902, PhysRevA.97.052315, PhysRevLett.128.170503, PhysRevLett.129.130502, PhysRevApplied.15.064020, PhysRevA.95.063849,PhysRevApplied.18.024076, XiaPhysRevLett.116.023601, PhysRevA.101.032329, PhysRevA.106.063707, PhysRevLett.128.203602, PhysRevA.99.043833,PhysRevA.90.043802, PhysRevX.3.031013,PhysRevA.97.062318, PhysRevA.104.053707, PhysRevResearch.4.033083} and experimentally~\cite{daiss2021quantum, reiserer2014quantum, hacker2016photon, PhysRevX.8.011018}, but are limited thus far to low-dimensional cases. Theoretically, high-dimensional photonic quantum systems also exhibit exotic fundamental physics regarding quantum nonlocality and Bell's theorem~\cite{erhard2018twisted, erhard2020advances,cozzolino2019high}. These are superior to low-dimensional systems, in improving the capacity of information processing and noise resilience~\cite{erhard2018twisted, erhard2020advances, cozzolino2019high, PhysRevX.8.041007, malikNatpho2016,mirhosseiniHighdimensionalQuantumCryptography2015,zhuHighdimensionalPhotonicEntanglement2021, Pirandola:20}, clock synchronization~\cite{tavakoli2015quantum}, and quantum metrology~\cite{fickler2012quantum}. These can also significantly simplify quantum circuit designs and enhance efficiencies in quantum computation~\cite{lanyon2009simplifying}.

The orbital angular momentum (OAM)~\cite{BLIOKH20151, bliokhSpinOrbitInteractions2015} is a useful resource for exploring high-dimensional quantum information techniques. By using bulk optics, such as spiral phase plates and parity sorters, a high-dimensional single-photon gate in an OAM-encoded basis was conducted experimentally~\cite{PhysRevLett.119.180510}. By fully utilizing the radial and azimuthal degrees of freedom of the photonic OAM, an \emph{equivalent} two-qubit controlled-NOT quantum gate has been demonstrated with a single photon encoded in four-dimensional (4D) OAM space in a recent experiment~\cite{optica.7.98}.  Although two-photon quantum gates between qubits were intensively studied, the counterpart in high-dimensional space is still elusive. We note that a multidimensional photon-photon gate has also been realized by using auxiliary photons and linear devices~\cite{jianweiNC13.1166}, but it is probabilistic.

A recent experiment has demonstrated that the $^{40}\textrm{Ca}^{+}$ ion has electrical quadrupole transitions and displays transition selection rules critically dependent on the spin angular momentum (SAM) and OAM of photons~\cite{nat.commun.7.12998}.

Inspired by this work~\cite{nat.commun.7.12998} and the scattering two-photon gate protocol~\cite{PhysRevLett.92.127902}, we theoretically propose a scheme based on the ion-cavity QED system to perform a two-photon quantum controlled phase-flip gate (CPF) with high fidelity by encoding two single photons in a 4D space spanned by photonic SAMs and OAMs.

This paper is organized as follows. In Sec.~\ref{sys}, we introduce the key idea and the basic system of our quantum gate and also present the quantum model for it. We explain a high-dimensional basis encoding in the $^{40}\textrm{Ca}^{+}$ ion, the scattering phase, and the six-step construction of the gate. Sec.~\ref{Exact numerical results} shows numerical simulation results of our gate performance, and evaluates in details the noise contributions to the gate infidelities. Sec.~\ref{implementation} discusses the practical system parameters for its experimental implementation. In the end, we conclude our findings in Sec.~\ref{conclusion}.

\section{System and model}\label{sys}
The ion-cavity QED system is depicted in Fig.~\ref{fig:FIG1}(a). A single $^{40}\textrm{Ca}^{+}$ ion is trapped in the center of a single-sided Fabry-P\'{e}rot cavity. Because the ion-cavity interaction is dependent on the SAM and OAM of the cavity mode, the system needs to be described by high-dimensional cavity QED (cQED).
 We focus on the electric quadrupole transition of $^{40}\textrm{Ca}^{+}$~\cite{nat.commun.7.12998}
 \begin{equation}
 |4^{2}S_{1/2}, m_J=\pm\frac{1}{2}\rangle \leftrightarrow |3^{2}D_{5/2},m^{'}_J=\pm \frac{1}{2}, \pm\frac{3}{2}, \pm\frac{5}{2}\rangle \;.
 \end{equation}
  We denote the two ground states
 \begin{equation}
  |4^{2}S_{1/2}, m_J=\pm\frac{1}{2}\rangle \equiv \{\mathinner{|\!\uparrow\rangle}, \mathinner{|\!\downarrow\rangle}\} \;, 
 \end{equation}
with frequency $\{\omega_{\downarrow}, \omega_{\uparrow}\}$, and the six excited magnetic sublevels as 
\begin{equation}
|m^{'}_J\rangle  \equiv |i_{\textrm{ion}}\rangle \;, 
\end{equation}
with $\ m^{'}_J\in \left\{-5/2, -3/2, -1/2, 1/2, 3/2, 5/2\right\}
$ corresponding to $i \in \{1_{\text{ion}}, 2_{\text{ion}}, 3_{\text{ion}}, 4_{\text{ion}}, 5_{\text{ion}}, 6_{\textrm{ion}}\}$, and $\omega_i$ for the frequency of excited state $| i_\text{ion} \rangle$, respectively.
 
We assume that the cavity modes with differential SAM ($s= \pm 1, 0$) and topological charges ($\ell = \pm 1, 0$) have a degenerate resonance frequency $\omega_c$. We neglect the intrinsic loss of the cavity. The cavity decay rate due to the input-output mirror is denoted by $\kappa$.
The two input single-photon pulses with frequency $\omega_p$ are encoded in their SAM and OAM, denoted as $|s, \ell\rangle$, and are successively injected to and reflected off the cavity. The input and reflected photons are separated via an optical circulator.

\subsection{Transition Selection Rules}
According to the transition selection rules of the $^{40}\textrm{Ca}^{+}$ ion, the quadrupole transitions require $\Delta m_J = \pm 2, \pm 1, 0$. Thus, there are $2 \times 5 = 10$ transitions involved. The ground state $\mathinner{|\!\downarrow\rangle}$ couples to $|j^{'}_{\text{ion}}\rangle$, with $j^{'} = 1_{\text{ion}},2_{\text{ion}},3_{\text{ion}},4_{\text{ion}},5_{\text{ion}}$ and $\mathinner{|\!\uparrow\rangle}$ couples to $|j^{''}_{\text{ion}}\rangle$ with $j^{''} = 2_{\text{ion}}, 3_{\text{ion}}, 4_{\text{ion}}, 5_{\text{ion}}, 6_{\text{ion}}$, see Fig.~\ref{fig:FIG1}(b). The coupling strength for the transition $|\!\downarrow\rangle \leftrightarrow \ket{j^{'}_{\text{ion}}}$ ($\mathinner{|\!\uparrow\rangle} \leftrightarrow \ket{j^{'}_{\text{ion}}}$) are $g_{j^{'}}$ ($g^{'}_{j^{''}}$). These are slightly different from each other with the multiplication of Clebsch-Gordan coefficients. We distinguish them in numerical simulations~\cite{PhysRevLett.119.253203}. Here, we assume they are identical and equal to $g$.

To select the $\mathinner{|\!\downarrow\rangle} \leftrightarrow \ket{5_{\text{ion}}}$ transition for our quantum gate, we apply a magnetic field $B$ to the ion.
The six magnetic sublevels are linearly separated in energy due to the Zeeman effect. The level energy is shifted by 
\begin{equation}
	\delta E=\mu_B g_D m^\prime_J B \;,
\end{equation}
where $g_{D}= 6/5$ is the Land\'e g-factor for the $D$ state. The ground states $\mathinner{|\!\downarrow\rangle}$ and $|\!\uparrow\rangle$ also split by 
\begin{equation}
\delta E=\mu_B g_S m_J B \;,
\end{equation}
where $g_{S}=2$ is the g-factor of the $S$ state, $\mu_{B}$ is the Bohr magneton and $\mu_{B}=14~\mega\hertz\cdot\milli\tesla^{-1}$.We denote the detuning between the adjacent excited magnetic sublevels as $\Delta=g_D\mu_B B$, and the detuning of the $\mathinner{|\!\downarrow\rangle} \leftrightarrow \ket{5_{\text{ion}}}$ and $\mathinner{|\!\uparrow\rangle} \leftrightarrow \ket{6_{\text{ion}}}$ transitions as
\begin{equation}
	\delta_{\Delta m_J = 2}=(g_{S}-g_{D})\mu_{B}B \;.
\end{equation}

According to angular momentum conservation, transitions happen only when the photons carry a total angular momentum of
\begin{equation}
L \equiv s + \ell = \{-2, -1, 0, 1, 2 \} \;.
 \end{equation}
  But the $\Delta m_J=0$ transition involves degenerate two-cavity modes with $L =0$ because the ion can absorb a photon in either state $|+1,-1\rangle$ or $|-1,+1\rangle$. Thus, we consider the remaining four transitions and photon states encoded in the basis of the 4D SAM-OAM hybrid space 
  \begin{equation}
  \{|\Psi\rangle\}=\{|-2\rangle, |-1\rangle, |+1\rangle, |+2\rangle\} \;.
  \end{equation}
With this chiral 4D cQED system, we can create quantum phase correlations between two single photons reflected off the Fabry-P\'{e}rot cavity and thus perform a two-photon quantum phase-flip gate.

\subsection{High-dimensional two-photon quantum controlled phase-flip gate}
The key idea of performing the high-dimensional two-photon quantum CPF gate is depicted with the quantum circuit in Fig.~\ref{fig:FIG1}(c). To perform the gate, we need to first induce a $\pi$ phase shift, conditioned on the ion spin state $\mathinner{|\!\downarrow\rangle}$, to a specific high-dimensional state of the first single-photon pulse $p_1$. The following step repeats the first for a second single-photon pulse $p_2$. Then, the ion is measured to project the three-body entangling state of the two single photons and the ion to a two-photon state. In doing so, the quantum CPF gate is accomplished for two traveling single photons.

The crucial step for the quantum CPF gate is to create a $\pi$ phase difference between a selective photonic state with the high-dimensional cQED system and other states. This is achieved with a controlled-$\mathbf{\tilde{Z}}_d$ gate with dimension $d=4$.
In practice, we have four cavity modes, corresponding to $|s = \pm 1, \ell = \pm 1\rangle_{\text{cav}}$. 

In experiments, the splitting  of cavity modes with different $\ell$ is typically very small, and can be further suppressed around tens of $\kilo\hertz$ with appropriate choices of mirror curvatures ~\cite{weiActiveSortingOrbital2020}. Thus, without loss of generality, we assume that these cavity modes are degenerate. We also consider that only the ionic  $\mathinner{|\!\downarrow\rangle} \leftrightarrow \ket{5_{\text{ion}}}$ transition is resonant with the cavity modes and the incident photon, i.e. 
\begin{equation}
	\omega_p = \omega_c = \omega_5 - \omega_\downarrow \equiv \omega_{5\downarrow}\;.
\end{equation}
This resonance condition between two successive photons and the cavity mode is critical to the success of the gate operation. Significant detunings between the input photons and the ionic transitions can result in a decline in the gate fidelity.
Other transitions related to the $\mathinner{|\!\downarrow\rangle}$ and $\mathinner{|\!\uparrow\rangle}$ states are off resonance with the cavity. This selective driving can be obtained by shifting the ionic states with a magnetic field $B$.

\subsection{Reflection coefficients for the input photon states}
\begin{figure}
  \centering
  \includegraphics[width=0.8\linewidth]{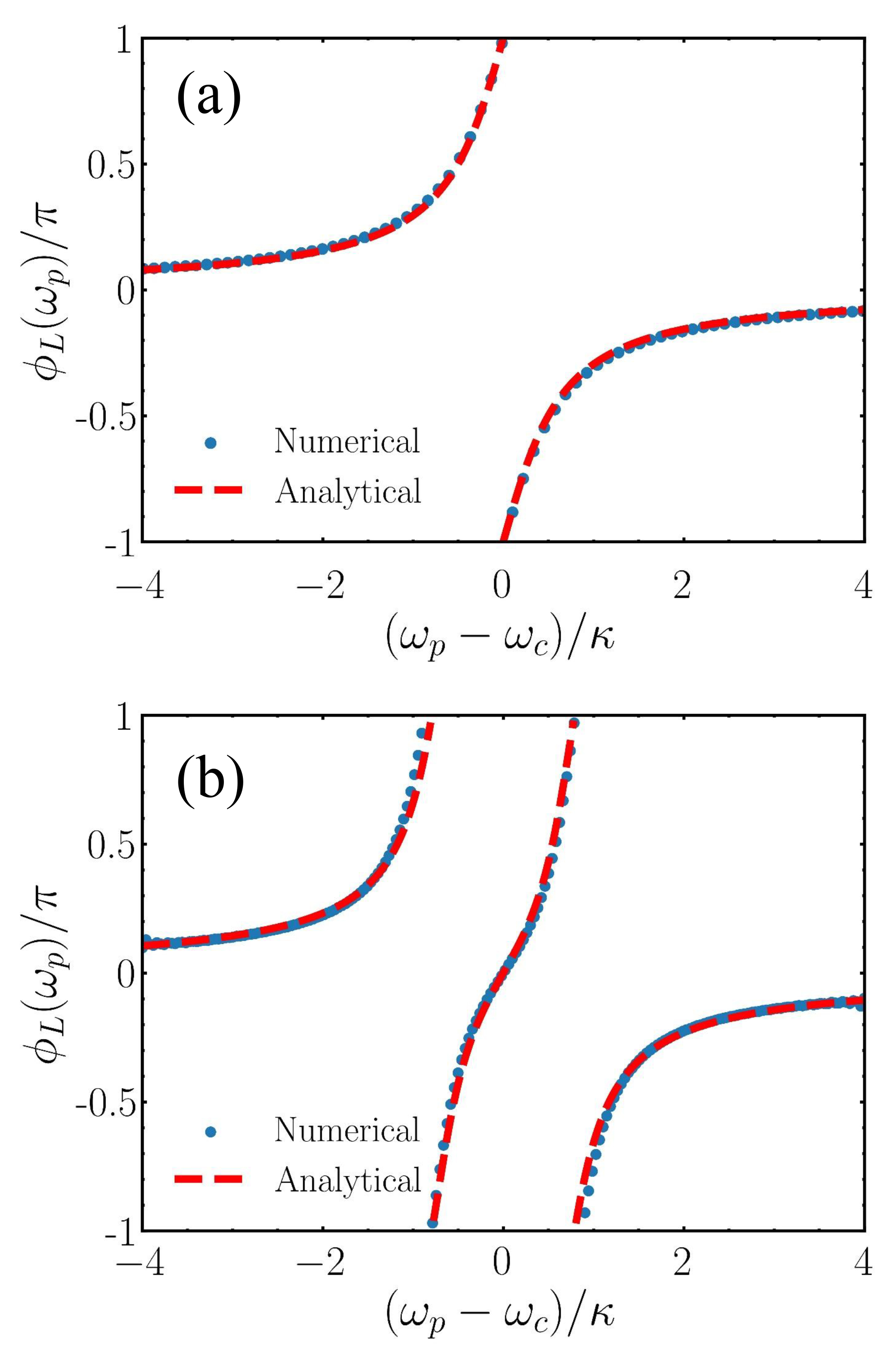} \\
  \caption{Numerical simulation (blue markers) and theoretical results (red dashed curves) of scattered photon phase shifts for the state  $\ket{2}$ versus detuning $\omega_p-\omega_c$. (a), the ion is initially prepared in the state other than $\mathinner{|\!\downarrow\rangle}$ (only show one case for example here). (b), the ion is in the $\mathinner{|\!\downarrow\rangle}$ state. Other parameters are $g/\kappa=3$ and $\Delta/\kappa=10$, which are experimentally accessible~\cite{hacker2016photon}.}
\label{fig:FIG2}
\end{figure}

The ion in state $\mathinner{|\!\uparrow\rangle}$ decouples from the cavity. In this case, the reflection coefficients for all input photonic states are equal and can be obtained by solving the Heisenberg equation of motion~\cite{PhysRevB.78.125318} as
\begin{equation}\label{eq:reflection2}
	r_0(\omega_p) = \frac{i(\omega_p-\omega_c)-\kappa}{i(\omega_p-\omega_c)+\kappa}.
\end{equation}
The phase shift on the input photon is shown by the red dashed curves in Fig.~\ref{fig:FIG2}(a). For an input single photon resonant with the cavity, $\omega_p = \omega_c$, we obtain $r_0(\omega_c) = -1$; i.e., all reflected photonic states acquire a global $\pi$ phase.
If the ion is in state $\mathinner{|\!\downarrow\rangle}$, the photonic states $\ket{-2}, \ket{-1}, \ket{1}$ still acquire a phase $\pi$, according to Eq.~(\ref{eq:reflection2}).

In contrast, the photonic state $\mathinner{|\!+2\rangle}$ couples to the cavity mode with $s = 1$ and $\ell =1$. This cavity mode  strongly interacts with the ionic transition  $\mathinner{|\!\downarrow\rangle} \leftrightarrow \ket{5_{\text{ion}}}$. Thus, the reflection coefficient of the photons is now given by
\begin{equation}\label{eq:reflection1}
	\begin{aligned}
		r(\omega_p)= \frac{(\omega_p-\omega_c+i\kappa)(\omega_p-\omega_{5\downarrow}+i\gamma)-g^2}{(\omega_p-\omega_c-i\kappa)(\omega_p-\omega_{5\downarrow}+i\gamma)-g^2}\;.
	\end{aligned}
\end{equation}
The reflected photon is subject to a phase shift $\phi_L(\omega_p)$. It is essentially different from the aforementioned detuned case due to the vacuum Rabi splitting of the cQED system. It is defined as 
\begin{equation}\label{eq:phiL}
\phi_L(\omega_p) = \text{Arg}[r(\omega_p)] \;,
\end{equation} 
when the ion is populated in the state $\mathinner{|\!\downarrow\rangle}$. Otherwise, it is calculated as 
\begin{equation}\label{eq:phiL}
\phi_L(\omega_p) = \text{Arg}[r_0(\omega_p)] \;.
\end{equation}
This analytical phase shift is shown by the red dashed curves in Fig.~\ref{fig:FIG2}(b).
Under on-resonance condition, we have 
\begin{equation}
	r(\omega_c) = (g^2 + \kappa\gamma)/(g^2- \kappa\gamma) \approx 1\;.
\end{equation}
Here, we utilize the strong coupling condition $g^2\gg \kappa\gamma$. Neglecting the global phase $\pi$, the state $\ket{2}$ equivalently acquires a $\pi$ phase shift with respect to all other photonic states.

Thus, if we prepare the initial ion state in a coherent superposition $(\mathinner{|\!\downarrow\rangle} - \mathinner{|\!\uparrow\rangle})/\sqrt{2}$ state, after reflected off the cQED system, only the state $\mathinner{|\!\downarrow\rangle}|2\rangle$ is subject to a relative $\pi$ phase shift. This is exactly the high-dimensional ion-photon CPF gate $U_{\text{ap}} = \left(\mathbf{1}_{4}, 0; 0, \mathbf{\tilde{Z}}_4\right)$,
with $\mathbf{1}_{4}$ representing the 4D identity matrix, and
\begin{equation}
	\mathbf{\tilde{Z}}_4= \begin{pmatrix}
		1 & 0 & 0 & 0 \\
		0 & 1 & 0 & 0 \\
		0 & 0 & 1 & 0 \\
		0 & 0 & 0 & -1  \\
	\end{pmatrix}\;,
\end{equation}
in the basis $\{\mathinner{|\!-2\rangle}, \mathinner{|\!-1\rangle}, \mathinner{|1\rangle}, \mathinner{|2\rangle}\}$.

\subsection{Gate operations}
Now we define the notation of the initial state for the gate operation.
We consider that both input photons are resonant with the cavity so that $\omega_p = \omega_c$.
The initial state of the two-photon pulses can be written as the product of superposition states
\begin{equation}
	|p_1,p_2\rangle_\text{in} = \sum_{M,N}\alpha_{M}\beta_{N}|M,N\rangle \;,
\end{equation}
where $M, N\in\{-2,-1,1,2\}$, $\int \sum_M |\alpha_{M}(t)|^2 dt =1$ and $\int \sum_N |\beta_{N}(t)|^2 dt =1$. This state is defined by the $16$ complex time-dependent functions $\alpha_{M}(t)\beta_{N}(t)$. For simplicity, we use the compact notation~\cite{hacker2016photon}
\begin{equation}
|m, n\rangle \equiv \alpha_{M}\beta_{N}|M,N\rangle 
\end{equation}
with $m, n \in \{-2,-1,1,2\}$
The two-photon state can then be rewritten in terms of the resonant $|2\rangle$ state as
\begin{equation}
	|p_1p_2\rangle_\text{in} = \sum_{i,j\neq 2}|i,j\rangle+\sum_{k\neq 2}(|k, 2\rangle+|2, k\rangle) + |2, 2\rangle \;.
\end{equation}
Considering the initial ionic state $\mathinner{|\!\downarrow\rangle}$, the initial system state is then
\begin{equation}
	|\psi\rangle_\text{in} = \mathinner{|\!\downarrow\rangle}(\sum_{i,j\neq 2}|i,j\rangle+\sum_{k\neq 2}(|k, 2\rangle+|2, k\rangle) + |2, 2\rangle) \;.
\end{equation}

Next, we discuss the detailed construction of the high-dimensional two-photon CPF gate according to the quantum circuit schematically shown in Fig.~\ref{fig:FIG1}(c).  

The ion is first prepared  in the state $(\mathinner{|\!\downarrow\rangle} - \mathinner{|\!\uparrow\rangle})/\sqrt{2}$ with a $\pi/2$ microwave pulse~\cite{PhysRevA.91.042307}. The second step is to reflect the first photon state $|p_1\rangle$ off the cavity. This equivalently performs a 4D controlled-$\mathbf{\tilde{Z}}_4$ operation between the ion and the first photon. By neglecting the global phase $\pi$, it flips the sign of all states related to state $\mathinner{|\!\downarrow,p_1=2\rangle}$. The resultant collective state then becomes
\begin{equation}
	\begin{aligned}
			|\psi\rangle_2 = \frac{1}{\sqrt{2}}\mathinner{|\!\downarrow\rangle}(\sum_{i,j\neq 2}|i,j\rangle+\sum_{k\neq 2}(|k, 2\rangle-|2, k\rangle ) - |2, 2\rangle) \\
		- \frac{1}{\sqrt{2}}\mathinner{|\!\uparrow\rangle}(\sum_{i,j\neq 2}|i,j\rangle+\sum_{k\neq 2}(|k, 2\rangle +|2, k\rangle)+ |2, 2\rangle) \;.
	\end{aligned}
\end{equation}

The third step rotates the ion on the two ionic ground states with a $-\pi/2$ mw pulse.
The fourth step performs the controlled-$\mathbf{\tilde{Z}}_4$ gate operation on the ion and the second photon.
It converts the system state to
\begin{equation}
	|\psi\rangle_4 = \mathinner{|\!\uparrow\rangle} (\sum_{k\neq 2}|k, 2\rangle+ |2,2\rangle)
	+ \mathinner{|\!\downarrow\rangle} (\sum_{i,j\neq 2}|i,j\rangle - \sum_{k\neq 2}|k, 2\rangle) \;.
\end{equation}
        
Finally, we again apply a $\pi/2$ rotation to the ionic ground states and measure them. Upon detecting the ion in the $\mathinner{|\!\downarrow\rangle}$ state, an additional $\pi$ phase is imprinted on the state related to the $|p_1 = 2\rangle$, resulting in a $\pi$ phase flip on the states $(\sum_{k\neq 2}|2,k\rangle+|2, 2\rangle)$, while the photonic state remains unchanged upon detection of $\mathinner{|\!\uparrow\rangle}$. Experimentally, this operation can be realized with a fast temporal switch, which separates the fluorescence photon from the ion and the working photons and directs the former to the single-photon detector~\cite{hacker2016photon}. To operate repeatedly, we can wait for enough long time so that the ion returns to its initial state. Subsequently, the photon pulses are separate in time. After measurement, we obtain the final two-photon state
\begin{equation}
			|p_1p_2\rangle_\text{f} = \sum_{i,j\neq 2}|i,j\rangle-\sum_{k\neq 2}|k,2\rangle + \sum_{k\neq 2}|2, k\rangle + |2, 2\rangle \equiv |\psi_\text{ideal}\rangle \;.
\end{equation}
Without including the global phase, the final state is independent of the outcome of the ionic state detection. Hence, the total circuit acts as a high-dimensional two-photon CPF gate with a truth table describing a gate operation:
\begin{equation}
    \begin{aligned}
    &\sum_{i,j\neq 2}|i,j\rangle \rightarrow \sum_{i,j\neq 2}|i,j\rangle \;,      &\sum_{k\neq 2}|k, 2\rangle \rightarrow -\sum_{k\neq 2}|k,2\rangle \;, \\
    &\sum_{k\neq 2}|2, k\rangle \rightarrow \sum_{k\neq 2}|2, k\rangle \;, &|2, 2\rangle \rightarrow |2, 2\rangle \;.
    \end{aligned}
\end{equation}

\section{Exact numerical results}\label{Exact numerical results}

\subsection{Simulation method}
Above we have presented an analytical description for the ideal gate's operation.
To evaluate the gate performance, we numerically simulate the actual operations with a full Hamiltonian for comparison with the aforementioned theoretical analysis. The full Hamiltonian for the system is given by $H = H_{\text{c-i}} + H_{\text{ph}} + H_{\text{int}}$:
\begin{equation}
    \begin{aligned}
        &H_{\text{c-i}} = H_{\text{i}} + H_{\text{g}} + H_{\text{d}} \;,  \\
        &H_{\text{ph}} =  \sum_{p=1,2}\sum_{L}\int \mathrm{d} \omega_p \ \omega_p \ b_{p,L}^\dagger(\omega_p)b_{p, L}(\omega_p) \;, \\
        &H_{\text{int}} =  \sum_{p=1,2}\sum_{L}\int \mathrm{d} \omega_p \ \kappa_{p} \ (a^\dagger_L b_{p, L}(\omega_p) + b_{p, L}^{\dagger}(\omega_p) a_L) \;,
    \end{aligned}
    \label{eq:hamiltonian}
\end{equation}
where $H_{\text{c-i}}$ characterizes the cavity-ion interactions, $H_{\text{ph}}$ describes the propagating photon pulses in the frequency domain, and $H_{\text{int}}$ describes the cavity-photon interactions. The annihilation operator for the cavity mode supporting total angular momentum $L$ is denoted as $a_L$, and
$b_{p, L}(\omega)$ is the annihilation operator for the $p$th photonic field with total angular momentum $L$ in the frequency domain. 
Here, we change to a reference frame rotating with the cavity frequency $\omega_c$.
We set $\omega_{\downarrow}$ as the reference energy. The ionic Hamiltonian in the rotating frame is
\begin{equation}
	H_{\text{i}} = \sum_{j=1}^{6} \Delta_{j}\sigma_{jj}
	+ \Delta_{\uparrow} \sigma_{\uparrow\uparrow} \;,
\end{equation}
with operators $\sigma_{jj}\equiv |j_{\text{ion}}\rangle\langle j_{\text{ion}}|$ and $\sigma_{\uparrow\uparrow} \equiv \mathinner{|\!\uparrow\rangle} \mathinner{\langle\uparrow\!|}$.  The detuning between the $j$-th excited magnetic sublevels and the cavity frequency is represented as $\Delta_{j}$. The coupling between each cavity mode $a_{L}$, $L \in \{-2, -1, 1, 2\}$, and the ions is described by 
\begin{equation}
	\begin{aligned}
		H_{\text{g}} = ( (g_1 a_{-2}\sigma_{1\downarrow}  + g^{'}_2  a_{-2}\sigma_{2\uparrow} + g_2 a_{-1} \sigma_{2\downarrow} + g^{'}_3 a_{-1} \sigma_{3\uparrow}\\
		 + g_4 a_1 \sigma_{4\downarrow} + g^{'}_5 a_{1} \sigma_{5\uparrow} + g_5 a_2 \sigma_{5\downarrow} + g^{'}_6 a_2 \sigma_{6\uparrow}) + \text{H.c.}) \;.
	\end{aligned}
\end{equation}
Here, the operator $\sigma_{j^{'}\downarrow} \equiv |j^{'}_{\text{ion}}\rangle \mathinner{\langle\downarrow\!|}$ denotes the transition $\mathinner{|\!\downarrow\rangle} \leftrightarrow |j^{'}_{\text{ion}}\rangle$ and $\sigma_{j^{''}\uparrow}$ for $\mathinner{|\!\uparrow\rangle} \leftrightarrow |j^{''}_{\text{ion}}\rangle$.
The driving Hamiltonian between two ground states is 
\begin{equation}
	H_{\text{d}} = \Omega(t) (\sigma_{\downarrow\uparrow}+ \text{H.c.}) \;,
\end{equation}
with microwave pulses $\Omega(t)\equiv \Omega_0 w(t)$, where $w(t)$ is the time-dependent box function (See Appendix.~\ref{Appendix_B}).  

The coupling between the cavity and different frequency modes of the photons $\kappa_p$ is assumed to be uniform. The nonuniform coupling $\kappa_p(\omega)$ introduces Lamb shifts to the dressed cavity resonance frequency. However, the Lamb shifts are very small, typically $\approx 0.01 \kappa$, and thus can be neglected, validating our assumptions~\cite{PhysRevA.91.063828, PhysRevApplied.17.054038}.
By expanding the Hamiltonian with the basis vectors Eq.~(\ref{basis}) in the low-excitation subspace, we obtain the discrete form of the Hamiltonian Eq.~(\ref{lisanham}) (For more details, see Appendix.\ref{Appendix_A}).

In simulations, we consider single-photon pulses
	\begin{equation}\label{photon_pulse}
		|\xi\rangle = \sum_{L} \int \mathrm{d}\omega_p f(\omega_p) b^{\dagger}_{p, L} (\omega_p) |\mathbf{0}\rangle \;,
	\end{equation}
	 where the normalized pulse-shape function $f(\omega_p)$ is Gaussian, 
	 \begin{equation}
	 	f(\omega_p) = \frac{1}{\sigma_\omega\sqrt{\pi}}\exp\left[-\frac{(\omega_p-\omega_c)^2}{\sigma_{\omega}^2}\right]\;,
	 \end{equation}
	with a central frequency $\omega_c$ and a bandwidth $\sigma_\omega$ for the inputs. These photons maximize the frequency bandwidth provided by the cavity $\sigma_{\omega} = \kappa$.

The analytic results for the phase shift of the reflected photons are confirmed by the full-Hamiltonian numerical simulations, see Fig.~\ref{fig:FIG2}, validating our idea for the high-dimensional two-photon quantum CPF gate.

\begin{figure}
	\centering
	\includegraphics[width=0.9\linewidth]{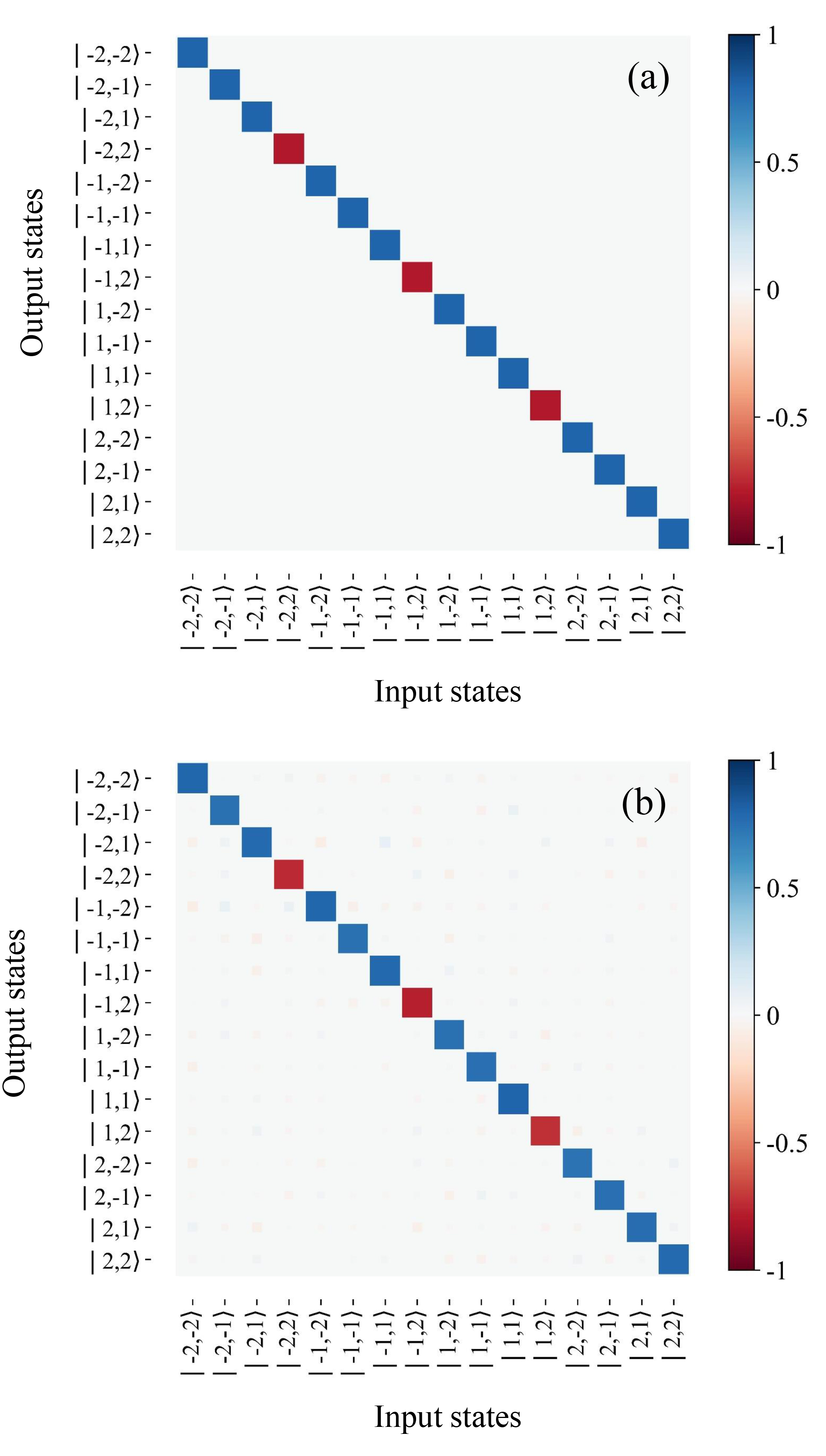} \\
	\caption{Truth table of the 4D two-photon CPF gate. (a) Ideal truth table. (b) Truth table with $\Delta/g=10$ in full-Hamiltonian simulations. The coupling strength and the Rabi frequency are $g/\kappa=3, \Omega_0/\kappa=5$.}
	\label{fig:FIG3}
\end{figure}

Now we clarify the evaluation of the output state and the gate-related fidelities. For an arbitrary input state $|\psi_\text{ph, in}\rangle  \equiv |p_1, p_2\rangle_\text{in}$ composed of two temporally separate identical single-photon pulses, we can solve the Schr\"{o}dinger equation and obtain the final photonic state after gate operations. Only considering the $\mathinner{|\!\downarrow\rangle} \leftrightarrow |5_{\text{ion}}\rangle$ transition in calculations, we obtain an ideal output $|\psi_\text{ideal}\rangle$. By including all $10$ possible transitions, the photon-photon gate output is $|\psi_\text{ph, out}\rangle$. Then, the fidelity of the output state is evaluated as 
\begin{equation}
	F (|p_1, p_2\rangle_\text{in}) = \left|\langle \psi_\text{ideal}| \psi_\text{ph, out}\rangle\right|^2 \;.
\end{equation}
To evaluate the performance of the quantum gate, we input $N = 16 \times 16 = 256$ initial two-photon states $|p_1p_2\rangle_{\text{in}}$ from the complete basis set $\mathcal{G}$: 
\begin{equation}
	\begin{aligned}
		 \mathcal{G} = &\left\{\frac{|0 \rangle + |1 \rangle}{\sqrt{2}},  \frac{|0 \rangle + i|1 \rangle}{\sqrt{2}}, \frac{|0 \rangle + |2 \rangle}{\sqrt{2}},\frac{|0 \rangle + i|2 \rangle}{\sqrt{2}}, |0\rangle,  |1\rangle, \right. \\
		& \left. \frac{|0 \rangle + |3 \rangle}{\sqrt{2}}, \frac{|0 \rangle + i|3 \rangle}{\sqrt{2}}, \frac{|1 \rangle + |2 \rangle}{\sqrt{2}}, \frac{|1 \rangle + i|2 \rangle}{\sqrt{2}},  |2\rangle, |3\rangle,\right. \\
		& \left. \frac{|1 \rangle + |3 \rangle}{\sqrt{2}}, \frac{|1 \rangle + i|3 \rangle}{\sqrt{2}},  \frac{|2 \rangle + |3 \rangle}{\sqrt{2}}, \frac{|2 \rangle + i|3 \rangle}{\sqrt{2}}  \right\}^{\otimes 2} \;.
	\end{aligned}
	\end{equation}
We then calculate the corresponding output states. The gate fidelity can be evaluated as
\begin{equation}
	F_\text{G} = \frac{1}{N}\sum_{|k\rangle \in \mathcal{G}} F(|k\rangle) \;,
\end{equation}
where $F(|k\rangle)$ is the state fidelity for the input two-photon state $|k\rangle$. Detailed simulation methods are provided in Appendix.~\ref{Appendix_B}.

\begin{figure*}
	\centering
	\includegraphics[width=1.0\linewidth]{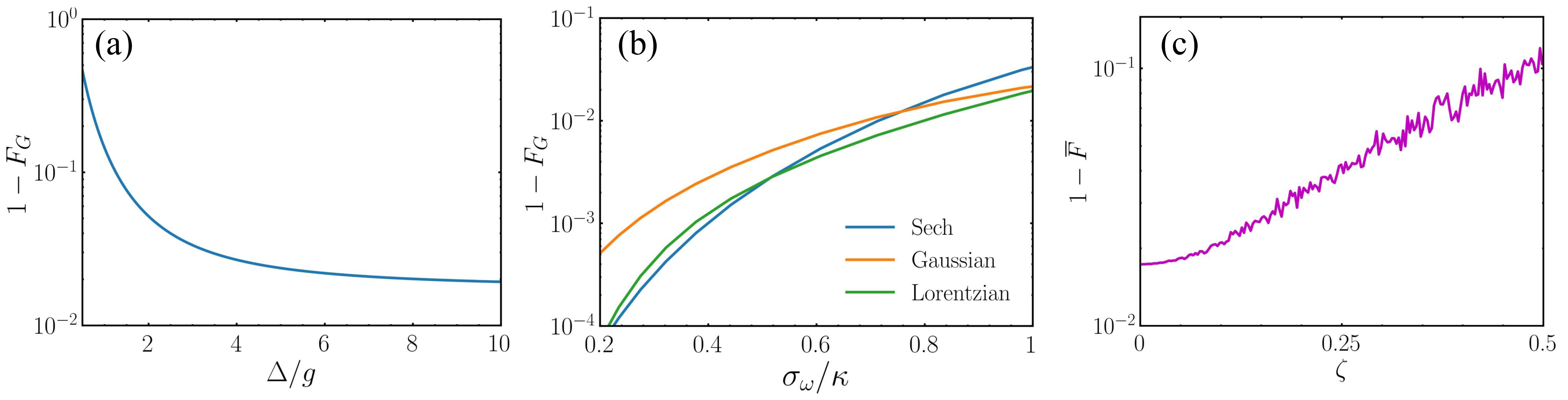} \\
	\caption{(a) Gate infidelity $1 - F_\text{G}$ versus the detuning-coupling ratio $\Delta/g$. (b) Gate infidelity $1 - F_\text{G}$ versus the incident pulse width $\sigma_\omega / \kappa$. We consider the three profiles of incident photons: Sech, Gaussian and Lorentzian. (c) Average gate infidelity $1 - \bar{F}$ versus Gaussian deviations in the control pulse area. The coupling strength and Rabi frequency in simulations are $g/\kappa=3$ and $\Omega_0/\kappa=5$, respectively.}
	\label{fig:FIG4}
\end{figure*}

\subsection{Truth Table}
Below we use the truth table to evaluate the performance of our 4D two-photon quantum gate. We input all $16$ pure photonic states $|i, j\rangle_\text{in}$, ($i, j \in \{-2, -1, 1, 2\}$, to the system. Figure~\ref{fig:FIG3}(a) shows the truth table for an ideal case. Then, we numerically calculate the final output state according to the quantum circuit with the full Hamiltonian in Eq.~(\ref{eq:hamiltonian}) for each input state.
When all $10$ transitions are included in the simulations, the truth table for a large detuning of $\Delta = 10 g$ is displayed in Fig.~\ref{fig:FIG3}(b). It is very close to the idea case. For each input state $|i, j\rangle_\text{in}$, we calculate the output state and the corresponding state fidelity $F (|i, j\rangle_\text{in})$. The average fidelity evaluated as $1/16 \sum_{|i, j\rangle_\text{in}} F (|i, j\rangle_\text{in})$ is high, reaching $99\%$, indicating a high success probability~\cite{Nature.619.495, PhysRevLett.129.130502,hacker2016photon}

\subsection{Noise Analysis}

\begin{table}[b]
	\caption{\label{table1}%
			Error contributions to the overall gate fidelity.}
		\begin{ruledtabular}
			\begin{tabular}{lr}
				\textrm{Source of gate errors}&
				\textrm{Error}\\
				\colrule
				Pulse shape distortion & $1.4\times 10^{-2}$ \\
				Transition to unwanted states & $0.2 \times 10^{-2}$ \\
				Cavity mode splitting &  $<1\times10^{-3}$\\
				Fluctuation of coupling strength $g$ & $<1\times10^{-3}$\\
				Fluctuations of control microwave pulse & $<1\times10^{-3}$\\
				Lamb shifts caused by inhomogeneous coupling & $< 1\times 10^{-5}$ \\
			\end{tabular}
		\end{ruledtabular}
	
\end{table}

\subsubsection{Fluctuating coupling strengths}
Next, we analyze the effects of different error contributions. The main results are summarized in Table~\ref{table1}. First, the trapped ions may not be well fixed within the cavity and experience a fluctuating coupling strength depending on its position $g(r)$. The gate fidelity, however, is robust against perturbations of the coupling strength $g$. This is because the vacuum Rabi splitting of two dressed modes protects the scattering phase factor from deviations, even if $g$ is reduced to a value comparable to the cavity decay $\kappa$. The contribution of a fluctuating coupling strength $g$ to the overall gate infidelity is of the order $10^{-4}$~\cite{PhysRevA.67.032305, PhysRevLett.92.127902}. 

\subsubsection{Detuning-coupling ratio}
The influence of the detuning-coupling ratio $\Delta/g$ on the gate fidelity is studied in Fig.~\ref{fig:FIG4}(a). As the ratio $\Delta/g$ increases from a vanishing value, the gate fidelity first increases rapidly and then becomes saturated. For a well accessible ratio $\Delta = 2 g$, the fidelity is already high, about $F_\text{G} \approx 95\%$, approaching saturation. When $\Delta =10 g$, the fidelity slightly improves to $98.4\%$. By using an experimentally available coupling strength $g \approx 2\pi \times 6 ~\mega\hertz$~\cite{PhysRevLett.124.013602} and a magnetic field  $B> 35~\milli\tesla$, we can obtain $\delta_{\Delta m_J = 2} \approx 2\pi \times 62.42~\mega\hertz > 10 g$. Therefore, we can perform a high-dimensional quantum gate with the ion-cavity system.

\subsubsection{Shapes and bandwidth of the incident photons}

Another major source of error arises from the distortion of photon pulses. In the most general case, the input single-photon state can be represented by Eq.~(\ref{photon_pulse}). After a sufficiently long time $t \gg \kappa^{-1}$, the output photon acquires a phase shift:
\begin{equation}
	|\xi(t)\rangle = \sum_{L} \int \mathrm{d}\omega_p f(\omega_p) e^{-i\omega_p t} e^{i\phi_{ L}(\omega_p)}b_{p, L}^\dagger(\omega_p) |\mathbf{0}\rangle \;.
\end{equation}
The first phase term $\exp(-i\omega_p t)$ represents the free evolution of the photon, while the second term $\exp[i\phi_{L}(\omega_p)]$ introduces a frequency- and angular-momentum-dependent scattering phase to the photon. Consequently, photon pulses with width $\sigma_\omega$ experience inhomogeneous scattering phases, deviating from the average scattering phase:
\begin{equation}
	\phi_{L}(\omega_p) \approx \phi_{L}(\omega_c) +\phi_{L}^{'}(\omega_c) (\omega_p-\omega_c) + \frac{\phi^{''}_{L}(\omega_c)}{2} (\omega_p - \omega_c)^2 \;.
\end{equation}

To investigate this distortion effect, we compare the real scattered photon $\ket{\xi(t)}=\exp(-iHt)\ket{\xi(0)}$ with an ideal photon that experiences no distortion, only delay, and acquires an average scattering phase $\phi_{L}(\omega_p)\approx \phi_{L}(\omega_c)$. The final average fidelity against pulse width is depicted in Fig.~\ref{fig:FIG4}(b). We observe that the gate infidelity $(1-F_\text{G})$ increases monotonically with the ratio $\sigma_\omega / \kappa$. Hence, to achieve low distortion and a good match of the scattering phase, the scattered photons must have a bandwidth $\sigma_\omega$ narrower than the cavity dissipation $\kappa$.
	
Furthermore, narrow-band photons generated from the trapped ions often deviate from Gaussian profiles. Thus, we explore the effect of pulse shapes on the gate fidelity, as shown in Fig.~\ref{fig:FIG4}(b). We compare Gaussian photons with Sech- and Lorentzian shaped photons, described by the following profiles:
	\begin{equation}
		f_{\text{S}}(\omega_p) = \sqrt{\frac{\pi}{2\sigma_\omega}} \text{Sech}\left[ \frac{\pi(\omega_p - \omega_c)}{\sigma_\omega}\right] \;,
	\end{equation}
	for the Sech profile and
	\begin{equation}
		f_{\text{L}}(\omega_p) = \frac{\sigma_\omega}{ \pi \left[(\omega_p - \omega_c)^2 + \sigma_\omega^2\right]} \;,
	\end{equation}
for the Lorentzian profile. We find that the exact pulse shape minimally affects gate transformations. The performance of the Gaussian pulse is marginally the same as the Lorentzian pulse when $\sigma_\omega = \kappa$, with $1 - F_{G} = 1.4 \%$ for the Gaussian profile, and $1.14\%$ for the Lorentzian profile. However, the fidelities for the Sech and Lorentzian pulses are higher than the Gaussian pulse, with infidelities $1 - F_{G} < 1\%$ when $\sigma_{\omega} > 0.75 \kappa$. Thus, near-unity fidelity of gate operation can be reached only if the narrow photon condition $\sigma_\omega \ll \kappa$ is satisfied. For a Gaussian wavepacket with bandwidth $\sigma_\omega = 0.2 \kappa$ and $\Delta / g = 10$, the gate fidelity reaches $F_{G} = 99.7\%$. The gate fidelity for the Lorentzian pulses under the same condition is $F_{G} > 99.9\%$, surpassing the lower threshold of quantum error correction~\cite{natphys4.463}. 

\subsubsection{Noise of microwave control pulses}
In practical operations, experimental imperfections can cause degradation of the gate operation. Here, the degradation mainly originates from the deviation of the control microwave pulses from the $\pi/2$ pulse area. We investigate this pulse area deviation on the average gate infidelity
\begin{equation}
	1 - \bar{F} = 1 - \frac{1}{N_r}\sum_{r=1}^{N_r} F_{G, r} \;.
\end{equation}
In each gate, we assume that the microwave pulses with amplitude $\Omega_0$ are subject to Gaussian noise with standard deviation $\zeta$. We investigate the gate fidelity averaged over $N_r = 50$ random gate operations versus the deviation strength $\zeta$, see Fig.~\ref{fig:FIG4}(c). Even for a deviation up to $\zeta \leq 0.25$, the average gate infidelity still remains relatively small, $1 - \bar{F} < 4\%$.  In the state-of-the-art experiment, the microwave control of trapped-ion qubits can be made very precise, with infidelities $ 1-\bar{F} \approx 10^{-4}-10^{-6}$, which correspond to very low $\zeta < 0.05$~\cite{PhysRevLett.113.220501, Bruzewicz2019}. Thus, the noise induced by the microwave control pulse has a small effect on the average gate infidelities. Clearly, this quantum gate is robust against the control imperfection.

\section{experimental implementation}\label{implementation}
Our system can be implemented by strongly coupling a trapped $^{40}\text{Ca}^{+}$ ion to a one-side Fabry-P\'erot microcavity, as demonstrated in \cite{PhysRevLett.124.013602, brandstatter2013integrated, PhysRevA.85.062308, PhysRevLett.110.043003}. One of the cavity mirrors has a relatively low reflectivity ($99.92\%$) as the output/input port, the other mirror has a relatively high reflectivity of $99.99\%$. Assuming a $600~\micro\meter$-long cavity, the total decay rate is estimated to be about $\kappa=2\pi\times 2~ \mega\hertz$. Photon pulses with bandwidth $\sigma_\omega = \kappa = 2\pi \times 2~\mega \hertz$ are sequentially reflected off the cavity~\cite{keller2004continuous,hacker2016photon, ding2013single}. The transition of the $E_2$ line $4^{2}S_{1/2}\leftrightarrow3^{2}D_{5/2}$ has a very long lifetime $\approx 1.045~\second$. Thus, the spontaneous decay rate $\gamma$ can be neglected. Using the experimentally available coupling strength $g=2\pi\times 6~\mega\hertz$~\cite{PhysRevLett.124.013602, nat.commun.7.12998, cui2018polarization, brandstatter2013integrated, PhysRevA.85.062308, PhysRevLett.110.043003} for the $\Delta m=2$ transition, the average gate fidelity can reach $\bar{F}>98\%$ when $B > 35~\milli\tesla$ and pulse noise deviation $\zeta < 0.1$. This performance is sufficiently high for many quantum information processing tasks.

\section{conclusion}\label{conclusion}
In summary, we have proposed the first deterministic high-dimensional two-photon quantum CPF gate by using the SAM- and OAM-dependent coupling between a ${}^{40}\text{Ca}^+$ and an optical cavity. The proposed gate achieves a high fidelity larger than $98\%$ and is robust against control imperfections. This approach can be extended to generate high-dimensional multiphoton entangled states, like cluster states and GHZ states, by adding auxiliary photons~\cite{PhysRevB.78.125318, PhysRevB.78.085307}.  Moreover, it can also make multinode quantum networks when the reflected photons are routed by polarization beam splitters. 
Therefore, this work opens an avenue for investigating fundamental physics of cQED systems in high-dimensional space and developing novel photonic quantum information techniques.

\section*{Acknowledgements}
This work was supported by the National Key R\&D Program of China (Grants No.~2019YFA0308700 and No.~2019YFA0308704), the National Natural Science Foundation of China (Grant No.~11890704), Innovation Program for Quantum Science and Technology (Grants No.~2021ZD0301400), the Program for Innovative Talents and Teams in Jiangsu (Grant No.~JSSCTD202138), China Postdoctoral Science Foundation (Grant No.~2023M731613), Jiangsu Funding Program for Excellent Postdoctoral Talent (Grant No.~2023ZB708). F. N. is supported in part by Nippon Telegraph and Telephone Corporation (NTT) Research, the Japan Science and Technology Agency (JST) [via the Quantum Leap Flagship Program (Q-LEAP), and the Moonshot R\&D Grant Number JPMJMS2061], the Asian Office of Aerospace Research and Development (AOARD) (via Grant No.~FA2386-20-1-4069), and the Office of Naval Research (ONR) Global (via Grant No.~N62909-23-1-2074). M.-Y. Chen thanks Y.-Z. Xiao for fruitful discussions. We thank the High Performance Computing Center of Nanjing University for allowing the numerical calculations on its blade cluster system.

\appendix
\begin{widetext}
\section{Hamiltonian Discretization}\label{Appendix_A}
To simulate the Hamiltonian Eq.~(\ref{eq:hamiltonian}) in the main text, we need to discretize $b_{p,L}(\omega)$ by introducing a finite but small frequency interval $\delta \omega = 2\sigma_\omega/N$ between two adjacent modes. To ensure that there is no significant change of results after the discretization, the frequency interval $\delta \omega$ should be chosen much smaller than the inverse of the gate operation time $T \approx 1 \micro \second$. The pulse width is chosen to be $\sigma_\omega = \kappa = 2~ \mega\hertz$~\cite{hacker2016photon}. We used $N=200$ for our simulation, which suffices because $\delta\omega \ll T^{-1}$. Then, the single-photon state becomes
\begin{equation}
	|\xi_p\rangle = \sum_{L, m} f_{m} b^{\dagger}_{p, L, m}|\mathbf{0}\rangle \;.
\end{equation}
Here, the pulse profile function is also discretized to 
\begin{equation}
f_{m}(\omega_m) = \frac{1}{\sigma_\omega\sqrt{\pi}}\exp[-(\omega_{m}-\omega_c)^2/\sigma_{\omega}^2] \;.
\end{equation}
The initial two-photon state is then represented as
\begin{equation}\label{p1p2}
	|\xi_{p_1p_2}\rangle_{\text{init}} = \sum_{L, m} \alpha_{L} f_{m} b^{\dagger}_{p_1, L, m} |\mathbf{0}\rangle \otimes \sum_{L^{'}, m^{'}} \beta_{L^{'}} f_{m^{'}} b^{\dagger}_{p_2, L^{'}, m^{'}}  |\mathbf{0}\rangle \equiv \sum_{L, L^{'}} \alpha_{L} \beta_{L^{'}} \ket{L, L^{'}} \equiv \ket{p_1, p_2}_{\text{init}} \;,	
\end{equation}
where $\alpha_{L}, \beta_{L^{'}}$ are normalized complex numbers. The equation Eq.~(\ref{p1p2})  corresponds to the compact notation $\ket{p_1p_2}_{\text{init}}$ of the two-photon state  in the main text.

After discretizing the basis states, we discretize the Hamiltonians $H_{\text{ph}}$ and $H_{\text{int}}$. Replacing $\int \omega \mathrm{d} \omega \rightarrow \sum_m \omega_m$, we have
\begin{equation}
	\begin{aligned}
		H_{\text{ph}} =   \sum_{p=1}^{2} \sum_{L\in\{-2, -1, 1, 2\}}\sum_{m=1}^{N} \omega_{m} b^{\dagger}_{p,L,m} b_{p, L, m} \;, \quad
		H_{\text{int}} = \sum_{p, L, m} \kappa_{p} (a_{L}^{\dagger} b_{p, L, m} + \text{\text{H.c.}})\;.
	\end{aligned}
\end{equation}

The ion-cavity system operates at cryogenic temperatures, thus thermal excitations can be neglected. Also, there is only one photon interacting with the ion-cavity system at each time, so we can study the Hamiltonian Eq.~(\ref{eq:hamiltonian}) in the subspace spanned by
\begin{equation}\label{basis}
	\begin{aligned}
		\ket{\Psi(t)} =& \sum_{L} a_{L}^{\dagger} (c_{1, L}(t) \ |0_L,\downarrow,0_{p_1},0_{p_2}\rangle + c_{2, L}(t) \ |0_L,\uparrow,0_{p_1},0_{p_2}\rangle) \\
		&+ \sum_{j^{'}=1,j^{'}\neq 3}^{5} p_{j^{'}}(t) \ \sigma_{j^{'},\downarrow} \ |0_L,\downarrow,0_{p_1},0_{p_2}\rangle + \sum_{j^{''}=2, j^{''}\neq 4}^{6} q_{j^{''}}(t) \ \sigma_{j^{''},\uparrow} \ |0_L,\uparrow,0_{p_1},0_{p_2}\rangle \\
		&+ \sum_{m, L} b^{\dagger}_{p_1, m,L} (\psi_{p_1 ,m,L,\downarrow}(t) \ |0_L,\downarrow,0_{p_1},0_{p_2}\rangle + \psi_{p_1,m,L,\uparrow}(t) \ |0_L,\uparrow,0_{p_1},0_{p_2}\rangle) \\
		&+ \sum_{m, L} b^{\dagger}_{p_2, m,L} (\phi_{p_2, m,L,\downarrow}(t) \ |0_L,\downarrow,0_{p_1},0_{p_2}\rangle + \phi_{p_2, m,L,\uparrow}(t) \ |0_L,\uparrow,0_{p_1},0_{p_2}\rangle) \;.
	\end{aligned}
\end{equation}
In this subspace, the Hamiltonian Eq.~(\ref{eq:hamiltonian}) is represented as a matrix form
\begin{equation}\label{lisanham}
		H = \begin{pmatrix}
			H_{\text{c-i}} & H_{\text{int}1} & H_{\text{int}2} \\[5pt]
			H^{T}_{\text{int}1} & H_{\text{ph}1}  & \mathbf{0}\\[5pt]
			H^{T}_{\text{int}2} & \mathbf{0} & H_{\text{ph}2} \\
		\end{pmatrix} \;.
\end{equation}
We now describe each Hamiltonian block in detail. The Hamiltonian in the upper left corner $H_{\text{c-i}}$ is a $16 \times 16$ matrix describing the ion-cavity interaction in the single-excitation subspace. Here we label $|n_{\text{cav},L}, j_{\text{ion}}, 0_{p_1}, 0_{p_2}\rangle \equiv |n_L, j\rangle$ for convenience. The ion-cavity Hamiltonian $H_{\text{c-i}}$ can be expressed as a combination of four block matrices
\begin{equation}
	H_{\text{c-i}} = 
	\begin{pmatrix}
		\mathbf{A} & \mathbf{B} \\
		\mathbf{B^{\dagger}} & \mathbf{D} \\
	\end{pmatrix} \;,
\end{equation}
where the block matrix $\mathbf{A}$ is 
\begin{equation}
	\mathbf{A} = \quad
	\begin{blockarray}{ccccccccc}
		\ket{1_{-2},\downarrow} & \ket{1_{-2},\uparrow} & \ket{1_{-1},\downarrow} & \ket{1_{-1},\uparrow} & \ket{1_{1},\downarrow} & \ket{1_{1},\uparrow} & \ket{1_{2},\downarrow} & \ket{1_{2},\uparrow} & \\
		\begin{block}{(cccccccc)c}
			0 & \Omega(t) & 0 & 0 & 0 & 0 & 0 & 0  &  \quad \bra{1_{-2},\downarrow}\\
			\Omega(t) & \omega_{5\uparrow} & 0 & 0 & 0 & 0 & 0 & 0 &  \quad \bra{1_{-2},\uparrow}\\
			0 & 0 & 0 & \Omega(t) & 0 & 0 &	0 & 0 & \quad \bra{1_{-1},\downarrow}\\
			0 & 0 & \Omega(t) & \omega_{5\uparrow} & 0 & 0 & 0 & 0 & \quad \bra{1_{-1},\uparrow}\\
			0 & 0 & 0 & 0 & 0 & \Omega(t) &	0 & 0 &  \quad \bra{1_{1},\downarrow}\\
			0 & 0 & 0 & 0 & \Omega(t) & \omega_{5\uparrow} & 0 & 0 &  \quad \bra{1_{1},\uparrow}\\
			0 & 0 & 0 & 0 & 0 & 0 &	0 & \Omega(t) & \quad \bra{1_{2},\downarrow}\\
			0 & 0 & 0 & 0 & 0 & 0 &	\Omega(t) & \omega_{5\uparrow} &  \quad \bra{1_{2},\uparrow}\\
		\end{block} 
	\end{blockarray}\;,
\end{equation}
the block $\mathbf{D}$ is 
\begin{equation}
	\mathbf{D} = \quad
	\begin{blockarray}{ccccccccc}
		\ket{0_{-2}, 1} & \ket{0_{-2}, 2} & \ket{0_{-1}, 2} & \ket{0_{-1}, 3} & \ket{0_{1}, 4} & \ket{0_{1}, 5} & \ket{0_{2}, 5} & \ket{0_{2}, 6} & \\
		\begin{block}{(cccccccc)c}
			\omega_{15} & 0 & 0 & 0 & 0 & 0 & 0 & 0 & \quad  \bra{0_{-2}, 1}\\
			0 & \omega_{25} & 0 & 0 & 0 & 0 & 0 & 0 & \quad  \bra{0_{-2}, 2}\\
			0 & 0 & \omega_{25} & 0 & 0 & 0 & 0 & 0 & \quad  \bra{0_{-1}, 2}\\
			0 & 0 & 0 & \omega_{35} & 0 & 0 & 0 & 0 & \quad  \bra{0_{-1}, 3}\\
			0 & 0 & 0 & 0 & \omega_{45} & 0 & 0 & 0 & \quad  \bra{0_{1}, 4}\\
			0 & 0 & 0 & 0 & 0 & 0 & 0 & 0 & \quad  \bra{0_{1}, 5}\\
			0 & 0 & 0 & 0 & 0 & 0 & 0 & 0 & \quad  \bra{0_{2}, 5}\\
			0 & 0 & 0 & 0 & 0 & 0 & 0 & \omega_{65} & \quad  \bra{0_{2}, 6}\\
		\end{block}
	\end{blockarray}\;,
\end{equation}
and the block $\mathbf{B}$ is
\begin{equation}
	\mathbf{B} = \quad 
	\begin{blockarray}{ccccccccc}
		\ket{0_{-2}, 1} & \ket{0_{-2}, 2} & \ket{0_{-1}, 2} & \ket{0_{-1}, 3} & \ket{0_{1}, 4} & \ket{0_{1}, 5} & \ket{0_{2}, 5} & \ket{0_{2}, 6} & \\
		\begin{block}{(cccccccc)c}
			g_1 & 0 & 0 & 0 & 0 & 0 & 0 & 0  &  \quad \bra{1_{-2},\downarrow}\\
			0 & g_2^{'} & 0 & 0 & 0 & 0 & 0 & 0  &  \quad \bra{1_{-2},\uparrow}\\
			0 & 0 & g_2 & 0 & 0 & 0 & 0 & 0  &  \quad \bra{1_{-1},\downarrow}\\
			0 & 0 & 0 & g_3^{'} & 0 & 0 & 0 & 0  &  \quad \bra{1_{-1},\uparrow}\\
			0 & 0 & 0 & 0 & g_4 & 0 & 0 & 0  &  \quad \bra{1_{1},\downarrow}\\
			0 & 0 & 0 & 0 & 0 & g_5^{'} & 0 & 0  &  \quad \bra{1_{1},\uparrow}\\
			0 & 0 & 0 & 0 & 0 & 0 & g_5 & 0  &  \quad \bra{1_{2},\downarrow}\\
			0 & 0 & 0 & 0 & 0 & 0 & 0 & g_6  &  \quad \bra{1_{2},\uparrow}\\
		\end{block}
	\end{blockarray} \;.
\end{equation}
The single-photon Hamiltonian $H_{\text{ph}1}$ can be written as a $8N \times 8N$ matrix. For simplicity, we encode the basis vectors as 
\begin{equation}
	\begin{alignedat}{2}
		\ket{1} &= \ket{0_{-2}, \downarrow, -2_{p_1}, 0_{p_2}} \;, &\quad
		\ket{2} &= \ket{0_{-2}, \uparrow, -2_{p_1}, 0_{p_2}} \;, \\
		\ket{3} &= \ket{0_{-1}, \downarrow, -1_{p_1}, 0_{p_2}} \;, &\quad
		\ket{4} &= \ket{0_{-1}, \uparrow, -1_{p_1}, 0_{p_2}}\;, \\
		\ket{5} &= \ket{0_{1}, \downarrow, 1_{p_1}, 0_{p_2}} \;, &\quad
		\ket{6} &= \ket{0_{1}, \uparrow, 1_{p_1}, 0_{p_2}} \;, \\
		\ket{7} &= \ket{0_{2}, \downarrow, 2_{p_1}, 0_{p_2}} \;, &\quad
		\ket{8} &= \ket{0_{2}, \uparrow, 2_{p_1}, 0_{p_2}}\;.
	\end{alignedat}
\end{equation}

The symbol $\mathbf{0}$ denotes the $N \times N$ zero matrix, $\bm{\tilde{\omega}}$ describes the $N \times N$ discretized eigenfrequency matrix for one single photon,  and $\bm{\tilde{\Omega}}$ is the driving term. These two matrices $\bm{\tilde{\omega}}$ and $\bm{\tilde{\Omega}}$ can be written in the form 
\begin{equation}
	\bm{\tilde{\omega}} = \quad
	\begin{blockarray}{cccc}
		\begin{block}{(ccc)c}
			\omega_1 &  &  &  \\
			& \ddots &  &  \\
			&  & \omega_N &  \\
		\end{block}
	\end{blockarray} \;, \qquad
	\bm{\tilde{\Omega}} = \quad
	\begin{blockarray}{cccc}
		\begin{block}{(ccc)c}
			\Omega(t) & & &  \\
			& \ddots & & \\
			& & \Omega(t) & \\
		\end{block}
	\end{blockarray} \;.
\end{equation}
We can then write the matrix elements explicitly as
\begin{equation}
	H_{\text{ph}1} = \quad
	\begin{blockarray}{ccccccccc}
		\ket{1} &\ket{2} &\ket{3} &\ket{4} &\ket{5} &\ket{6} &\ket{7} &\ket{8} \\
		\begin{block}{(cccccccc)c}
			\bm{\tilde{\omega}} & \bm{\tilde{\Omega}} & \mathbf{0} & \mathbf{0} &\mathbf{0}  &\mathbf{0} &\mathbf{0}& \mathbf{0}&  \quad \bra{1} \\
			\bm{\tilde{\Omega}} & \bm{\tilde{\omega}} & \mathbf{0} & \mathbf{0} &\mathbf{0} & \mathbf{0}& \mathbf{0} & \mathbf{0}&  \quad \bra{2}\\
			\mathbf{0} & \mathbf{0} & \bm{\tilde{\omega}} & \bm{\tilde{\Omega}} & \mathbf{0} & \mathbf{0} & \mathbf{0} & \mathbf{0} &  \quad \bra{3} \\
			\mathbf{0}& \mathbf{0} & \bm{\tilde{\Omega}} & \bm{\tilde{\omega}} & \mathbf{0} & \mathbf{0} & \mathbf{0} & \mathbf{0} &  \quad \bra{4} \\
			\mathbf{0} & \mathbf{0} & \mathbf{0} & \mathbf{0} & \bm{\tilde{\omega}} & \bm{\tilde{\Omega}} & \mathbf{0} & \mathbf{0} &  \quad \bra{5} \\
			\mathbf{0}& \mathbf{0} & \mathbf{0} & \mathbf{0} & \bm{\tilde{\Omega}} & \bm{\tilde{\omega}} & \mathbf{0} & \mathbf{0} &  \quad \bra{6} \\
			\mathbf{0}& \mathbf{0} & \mathbf{0} & \mathbf{0} & \mathbf{0} & \mathbf{0} & \bm{\tilde{\omega}} & \bm{\tilde{\Omega}} &  \quad \bra{7} \\
			\mathbf{0}& \mathbf{0} & \mathbf{0} & \mathbf{0} & \mathbf{0} & \mathbf{0} & \bm{\tilde{\Omega}} & \bm{\tilde{\omega}} &  \quad \bra{8} \\
		\end{block}
	\end{blockarray} \;.
\end{equation}
The Hamiltonian of the second photon $H_{\text{ph}2}$ is of the same structure. For the interaction Hamiltonian of the first photon and ion-cavity system $H_{\text{int}1}$, the matrix elements are
\begin{equation}
	H_{\text{int}1} = \\
	\begin{blockarray}{ccccccccc}
		\ket{1} &\ket{2} &\ket{3} &\ket{4} &\ket{5} &\ket{6} &\ket{7} &\ket{8} \\
		\begin{block}{(cccccccc)c}
			\bm{\tilde{\kappa}_{1}(t)} & \bm{\tilde{\kappa}_{1}(t)} & \mathbf{0} & \mathbf{0}  & \mathbf{0}   & \mathbf{0}   & \mathbf{0}  & \mathbf{0} &  \quad \bra{1_{-2}, \downarrow, 0_{p_1}, 0_{p_2}}\\
			\bm{\tilde{\kappa}_{1}(t)} & \bm{\tilde{\kappa}_{1}(t)} & \mathbf{0} & \mathbf{0}  & \mathbf{0}  & \mathbf{0}  & \mathbf{0} & \mathbf{0} &   \quad \bra{1_{-2}, \uparrow, 0_{p_1}, 0_{p_2}}\\
			\mathbf{0} & \mathbf{0} & \bm{\tilde{\kappa}_{1}(t)} & \bm{\tilde{\kappa}_{1}(t)} & \mathbf{0} & \mathbf{0}  & \mathbf{0}  & \mathbf{0} & \quad  \bra{1_{-1}, \downarrow, 0_{p_1}, 0_{p_2}}\\
			\mathbf{0}  & \mathbf{0}  & \bm{\tilde{\kappa}_{1}(t)} & \bm{\tilde{\kappa}_{1}(t)} & \mathbf{0} & \mathbf{0} & \mathbf{0} & \mathbf{0} & \quad  \bra{1_{-1}, \uparrow, 0_{p_1}, 0_{p_2}}\\
			\mathbf{0} &  \mathbf{0} & \mathbf{0} & \mathbf{0} & \bm{\tilde{\kappa}_{1}(t)} & \bm{\tilde{\kappa}_{1}(t)} & \mathbf{0} & \mathbf{0} & \quad \bra{1_{1}, \downarrow, 0_{p_1}, 0_{p_2}}\\
			\mathbf{0} & \mathbf{0}  & \mathbf{0}  & \mathbf{0}  & \bm{\tilde{\kappa}_{1}(t)} & \bm{\tilde{\kappa}_{1}(t)} & \mathbf{0} & \mathbf{0} & \quad  \bra{1_{1}, \uparrow, 0_{p_1}, 0_{p_2}}\\
			\mathbf{0} & \mathbf{0}  & \mathbf{0}  & \mathbf{0}  & \mathbf{0}  & \mathbf{0} & \bm{\tilde{\kappa}_{1}(t)} & \bm{\tilde{\kappa}_{1}(t)} & \quad  \bra{1_{2}, \downarrow, 0_{p_1}, 0_{p_2}}\\
			\mathbf{0} & \mathbf{0}  & \mathbf{0}  & \mathbf{0}  & \mathbf{0}  & \mathbf{0} & \bm{\tilde{\kappa}_{1}(t)} & \bm{\tilde{\kappa}_{1}(t)} & \quad  \bra{1_{2}, \uparrow, 0_{p_1}, 0_{p_2}} \\
			\mathbf{0} & \mathbf{0}  & \mathbf{0}  & \mathbf{0}  & \mathbf{0}  & \mathbf{0}  & \mathbf{0}  & \mathbf{0} & \quad\bra{0_{-2}, 1, 0_{p_1}, 0_{p_2}} \\
			\vdots & \vdots & \vdots & \vdots & \vdots & \vdots & \vdots & \vdots &  \quad\bra{0_{-2}, 2, 0_{p_1}, 0_{p_2}}\\
			&  &  &  &  &  &  &  &  \quad\bra{0_{-1}, 2, 0_{p_1}, 0_{p_2}}\\
			&  &  &  &  &  &  &  &  \quad\bra{0_{-1}, 3, 0_{p_1}, 0_{p_2}}\\
			&  &  &  &  &  &  &  &  \quad\bra{0_{1}, 4, 0_{p_1}, 0_{p_2}}\\
			&  &  &  &  &  &  &  &  \quad\bra{0_{1}, 5, 0_{p_1}, 0_{p_2}}\\
			\vdots & \vdots & \vdots & \vdots & \vdots & \vdots & \vdots & \vdots &  \quad\bra{0_{2}, 5, 0_{p_1}, 0_{p_2}}\\
			\mathbf{0} & \mathbf{0} & \mathbf{0} & \mathbf{0} & \mathbf{0} & \mathbf{0} & \mathbf{0} & \mathbf{0} &  \quad\bra{0_{2}, 6, 0_{p_1}, 0_{p_2}}\\	
		\end{block}
	\end{blockarray}\;. \qquad
\end{equation}
Here, $\bm{\tilde{\kappa}_{1}(t)} = \left[ \kappa_1(t), \kappa_1(t) \cdots \kappa_1(t) \right]$ is a $1 \times N$ row vector. The Hamiltonian of the second photon $H_{\text{int}2}$ has the same form as $H_{\text{int}1}$, and only requires the substitution of the corresponding elements $\kappa_1(t) \rightarrow \kappa_2(t)$ and basis vectors $\ket{0_L, j, L_{p_1}, 0_{p_2}} \rightarrow \ket{0_L, j, 0_{p_1}, L_{p_2}}$.
\medskip
\section{Simulation method}\label{Appendix_B}
Our goal is to simulate the final output two-photon state after gate operations. To achieve this goal, we use the Trotter-Suzuki formula, which is a more computationally-efficient approach to directly compute the time evolution of the given initial two-photon state $\ket{\xi_{p_1p_2}(T)} = U(T)\ket{\xi_{p_1p_2}(0)}$.
Here, $U(T) = \exp(iHT)$ is the time-evolution operator satisfying $U(t, t_0) = U(t, t_i)U(t_i, t_0)$. Thus, we can expand the time-evolution operator as $U(T) = U(T, T-\Delta t) U(T-\Delta t, T-2\Delta t)\cdots U(\Delta t, 0)$. The Trotter-Suzuki formula states that for a general Hamiltonian $H = H_1 + H_2$, with two non-commuting parts $[H_1, H_2]\neq 0$, the time evolution operator can be approximated as 
\begin{equation}
	U(\Delta t) = \exp(-iH\Delta t) = \exp(-iH_1 \Delta t) \exp(-i H_2 \Delta t) \exp(-i (\Delta t)^2 [H_1, H_2]) \approx  \exp(-iH_1 \Delta t) \exp(-i H_2 \Delta t) \;.
\end{equation} 
For an infinitesimal time interval $\Delta t$, the error is negligible. More generally, for $H = \sum_{\alpha=1}^{N_H} H_\alpha$, the time-evolution operator can be expressed as 
\begin{equation}\label{Trotter-Suzuki}
	U(T) = \prod_{n=1}^{N} \prod_{\alpha=1}^{N_H} \exp({-i H_\alpha T/n}).
\end{equation}
We use this general Trotter-Suzuki formula Eq.~(\ref{Trotter-Suzuki}) to simulate the high-dimensional two-photon CPF gate operations according to Fig.~1(c) in the main text. 
The only time-dependent elements in the Hamiltonian Eq.~(\ref{lisanham}) are $\Omega(t), \kappa_1(t)$ and $\kappa_2(t)$. These are the control parameters for different gate operations in Fig.~1(c) in the main text. To be more precise, we divide the time interval $[0, T]$ into six parts $t_i, i=1,2,3,4,5,6$, where $[t_{i-1}, t_i]$ denotes the time interval of the $i$-th gate operation. The controlled microwave pulse $\Omega(t)=\Omega_0 w(t)$ is a segmented function
\begin{equation}
	\Omega(t) = \left \{
	\begin{array}{rcl}
		&\Omega_0  & \quad 0 \leq t \leq t_1 \ \& \  t_4  \leq t \leq t_5\\
		&\Omega_0 \exp(i\pi) & \quad t_2 \leq t \leq t_3 \\
		& 0  & \quad \text{others}\\
	\end{array}
	\right. \;.
\end{equation} 
Here, $t_1 = t_5 - t_4 = \pi/(4\Omega_0)$, which ensures that the pulse area is $\pi/2$. The piecewise function $\Omega(t)$ corresponds to $\pi/2, -\pi/2, \pi/2$ rotations to the ion shown in Fig.~1(c) in the main text. To simulate the two ion-photon controlled-$\tilde{Z}_4$ gate, we set the  two coupling strengths $\kappa_1(t)$ and $\kappa_2(t)$ as
\begin{equation}
	\kappa_1(t) = \left \{
	\begin{array}{rcl}
		& \kappa_1 & \quad t_1 \leq t \leq t_2 \\
		& 0 & \quad \text{others} \\
	\end{array}
	\right. \;, \qquad
	\kappa_2(t) = \left\{
	\begin{array}{rcl}
		& \kappa_2 & \quad t_3 \leq t \leq t_4 \\
		& 0 & \quad \text{others} \\
	\end{array}
	\right. \;.
\end{equation}
Here, we set $t_2 - t_1 = t_4 - t_3 = 10\kappa^{-1}$ in order to ensure that the photons are completely scattered off the cavity. \vspace{0.2cm}

To summarize, the simulation procedure is as follows 
\begin{enumerate}
	\item Prepare the initial state $\ket{\Psi(t=0)} = \ket{0_{\text{cav}}, \downarrow} \otimes |\xi_{p_1p_2}(0)\rangle$ according to Eq.~(\ref{p1p2}). 
	
	\item Time-evolve the system $	\ket{\Psi(T)} = \exp(-iH(\Omega, \kappa_1, \kappa_2)(t)T) \ \ket{\Psi(0)}.$
	
	\item Measure the ionic state and trace over the cavity degrees of freedom to obtain the final two-photon state  $|\xi_{p_1p_2}(T)\rangle$.
	
	\item Compare the simulated $|\xi_{p_1p_2}(T)\rangle$ with the ideal two-photon state $|\tilde{\xi}_{p_1p_2}(T)\rangle$, which experiences no distortion, and acquires an average scattering phase $\phi_{L}(\omega_p) \approx \phi_{L}(\omega_c)$ in each scattering process. Then, the output state fidelity is obtained via $F = | \langle \xi_{p_1p_2}(T)| \tilde{\xi}_{p_1p_2}(T)\rangle|^2$. 
	
	\item Repeat the above four procedures for $N$ input states and compute the gate fidelity $F_G = \frac{1}{N} \sum_{n=1}^{N} F_n$.
\end{enumerate}

\section{Discussion on the coupling strength $g$}
For ionic states $\{ \ket{S, m_S}, \ket{D, m_D}\}$, the cavity couples with ionic states with different coupling strengths. We assume $g_{j^{'}=3} = g_{j^{''}=4} \equiv g$ and the vacuum coupling strength is $g_0$. The coupling strength for the transition $\ket{S, m_S} \leftrightarrow \ket{D, m_D}$ is $g_{j^{'}} = C(J_S m_S,2q; J_D m_D)g_0$, where $C(J_S m_S,2q; J_D m_D)$ is the Clebsch-Gordan coefficient given by a Wigner 3-j symbol~\cite{PhysRevLett.119.253203}.
\begin{equation}
	C(J_Sm_S,2q; J_D m_D) = (-1)^{J_S-2+m_D} \sqrt{2J_D+1} \left( \begin{array}{ccc}
		J_S & 2 & J_D \\
		m_S & q & -m_D \\
	\end{array} \right )
\end{equation}
We estimate that $g_{j^{'}=1} = 1/\sqrt{6} g$, $g_{j^{'}=2} = \sqrt{2/3} g$,  $g_{j^{'}=4} = \sqrt{3} /2 g$, $g_{j^{'}=5} = \sqrt{3/5} g$, and $g_{j^{'}} = g^{'}_{j^{''}}$ in our simulation.   

\end{widetext}

\end{document}